\documentclass{aa}  

\emergencystretch=\maxdimen     
\usepackage{graphicx}

\usepackage{xcolor}

\usepackage{txfonts}

\usepackage{subcaption}
\usepackage{hyperref}
\usepackage{amsmath}
\usepackage[]{algorithm2e}

\begin{document} 

   \title{An improved algorithm for separating clock delays from ionospheric effects in radio astronomy}
   \author{C.M. Cordun
          \inst{1,2}
          \and
          M.A. Brentjens \inst{1}
          \and H.K. Vedantham\inst{1,2}
          \and M. Mevius \inst{1}
          }

   \institute{ASTRON, Netherlands Institute for Radio Astronomy, Oude Hoogeveensedijk 4, Dwingeloo, 7991PD, The Netherlands\\
              \email{cordun@astron.nl}
         \and
             Kapteyn Astronomical Institute, University of Groningen, PO Box 800, 9700 AV, Groningen, The Netherlands
        }
        
   \date{Received September 15, 1996; accepted March 16, 1997}

  \abstract{Low-frequency radio observations are heavily impacted by the ionosphere, where dispersive delays can outpace even instrumental clock offsets, posing a serious calibration challenge. Especially below 100MHz, phase unwrapping difficulties and higher-order dispersion effects can complicate the separation of ionospheric and clock delays.}
{We address this challenge by introducing a method for reliably separating clock delays from ionospheric effects, even under mediocre to poor ionospheric conditions encountered near solar maximum.}
{The approach employs a key technique: we modelled our likelihood space using the circular Gaussian distribution (von Mises random variable) rather than non-circular distributions that suffer from 2$\pi$ phase ambiguities. This ensures that noisier data are weighted less heavily than cleaner data during the fitting process.}
{The method reliably separates clock delays and ionospheric terms that vary smoothly in time whilst providing a good fit to the data. A comparison with the clock–ionosphere separation approach used in standard LOFAR data processing shows that our technique achieves significant improvements. In contrast to the old algorithm, which often fails to return reliable results below 100MHz even under good ionospheric conditions, the new algorithm consistently provides reliable solutions across a wider range of conditions.}
{This new algorithm represents a significant advance for large-scale surveys, offering a more dependable way to study ionospheric effects and furthering research in ionospheric science and low-frequency radio astronomy.}


   \maketitle
%

\section{Introduction}

The new generation of low-frequency ($\lesssim 1{\rm GHz}$) radio interferometers is significantly improving our understanding of galaxies and galaxy clusters, both at low and high redshifts \citep{de2017gentle, saxena2018discovery, hoang2018radio, reis2020high}, active galactic nuclei \citep{brienza2016lofar, harwood2016fr}, and stellar systems \citep{villadsen2019ultra, climent2020milliarcsecond, callingham2021population}, among others. More recently, significant efforts have been made to explore the lowest frequencies observable from Earth (down to $\sim 10,{\rm MHz}$), known as the decametre regime. These studies have not only deepened our knowledge of traditional areas accessed by radio astronomy but have also opened up new possibilities in the fields of exoplanetary science \citep{zarka2007plasma, lynch2018detectability, kavanagh2023hunting} and the early Universe \citep{gehlot2019first,gehlot2020aartfaac}.

The Low-Frequency Array \citep[LOFAR;][]{VanHaarlem2013} is one of the leading instruments for observing the sky in the decametre regime, alongside the Owens Valley Long Wavelength Array \citep[OV-LWA;][]{eastwood201921} and the New Extension in Nançay Upgrading LOFAR \citep[NenuFAR;][]{zarka2015nenufar}. Even with next-generation facilities such as the Square Kilometre Array \citep[SKA;][]{hall2008square} coming online, LOFAR remains unique due to its combination of high sensitivity, fine resolution, and ability to observe at frequencies below $50~\text{MHz}$.

LOFAR is a distributed radio interferometer located in Europe, with the highest density of antenna elements in the Netherlands \citep{VanHaarlem2013}. The array includes 24 core stations within a 2 km radius, 14 remote stations spread across the Netherlands up to 100 km from the core, and 14 international stations, with two more under construction. Each station is equipped with two types of antennas: low-band antennas (LBAs), which function in the 10–90 MHz range, and high-band antennas (HBAs) for the 110–250 MHz range. For this study, we focused on the LBA system, as calibration methods for the HBA system are already well established.

LOFAR’s unique ability to probe the lowest observable frequencies at arcsecond resolutions can only be exploited if ionospheric propagation effects are properly calibrated and removed from the data. The same ionospheric calibration solutions also provide exceptional electron column density -- called total electron content (TEC) -- measurements, which are useful for geophysical research \citep{mevius2016probing, fallows2020lofar}. However, the ionospheric phase varies linearly with wavelength. The dynamic and direction-dependent nature of this phase has made ionospheric calibration at the lowest frequencies accessible to LOFAR a stubborn challenge. This typically precludes us from capturing thermal-noise-limited astronomical images \citep{de2018effect, de2019systematic}.

A specific challenge in LOFAR calibration is the so-called `clock--TEC separation’ problem. The problem involves separating systematic phase offsets caused by clock drifts between stations from those caused by dispersive ionospheric propagation. Station delays, which arise from instrumental differences, affect both the calibrator and the target field. In contrast, ionospheric effects are direction dependent and vary significantly between the calibrator and the target, especially if they are separated by more than a few degrees. The goal of clock--TEC separation is to transfer only the station delays from the calibrator to the target while excluding ionospheric contributions.

An algorithm for clock--TEC separation is implemented in the LOFAR \texttt{losoto} module \citep{de2019systematic}. While effective for LOFAR HBA datasets, this approach often fails when applied to LBA observations due to larger ionospheric effects, a larger relative bandwidth, and lower signal-to-noise per ionospheric coherence time. A larger relative bandwidth helps in separating clock effects, which are proportional to $\nu$, from TEC effects, which vary as $1/\nu$ (based on a first-order Taylor expansion), assuming the algorithm can handle the $2\pi$ phase ambiguities caused by the larger magnitude of the effects. This study is dedicated to finding an algorithm capable of handling these challenges.

As a workaround, the current practice involves transferring complete phase solutions from the calibrator to the target without separating station delays from ionospheric effects. While effective, this method limits the efficiency of large-scale surveys and constrains observing strategies, as it requires two simultaneous beams in the sky.

In this paper we present a new algorithm that reliably separates station delays from ionospheric effects, even in noisy, low-frequency data. The method is an improvement over the existing approach in two ways. First, it replaces traditional phase unwrapping with the use of the von Mises distribution, which is more robust to noise. Second, it uses a more complex likelihood-based fitting technique than least squares. Although this method is computationally more intensive, it accounts for noise levels in the data, resulting in more accurate results.

The paper is structured as follows. Section \ref{sec:obs} describes the observations used to test the algorithm and outlines the processing steps for obtaining phase solutions. The new algorithm is detailed in Sect. \ref{sec:method}, which is followed by the results (Sect. \ref{sec:results}) and a comparison with the standard software (Sect. \ref{sec:disc}). Finally, Sect. \ref{sec:conclusion} summarises the work and its broader implications.

\section{Observation and processing}\label{sec:obs}

The dataset that we used to test the algorithm was acquired with LOFAR’s LBA system in interferometric mode. The observation spans 4 hours, from 13.8MHz to 60.5MHz, and uses 35 of the 38 Dutch LOFAR stations in the LBA-sparse mode \citep{VanHaarlem2013}. The target — the Tau Boötis b field — and the calibrator — 3C196 — were observed simultaneously using LOFAR’s multi-beam capability. The observations were primarily scheduled at night to minimise the impact of radio-frequency interference (RFI) reflected off the ionosphere. However, continuous bright RFI below 20~MHz made the ultra-low-frequency data unusable. Further details about the observation are provided in \autoref{tab:obs}.

\begin{table}[!ht]
\centering
\caption{Details of the LOFAR array setup and the observational dataset.}
\label{tab:obs}
\begin{tabular}{c|c}
\hline\hline
Parameter              & Value \\
\hline
Instrument             & LOFAR-LBA \\
Array Setup            & LOFAR Remote \\
Number of Stations     & 35 \\
Antenna Configuration  & LBA Sparse \\
SAS ID                 & L2037672 \\
Minimum Frequency      & 13.8~MHz \\
Maximum Frequency      & 60.5~MHz \\
Subbands per Dataset   & 238 \\
Channels per Subband   & 256 \\
Frequency Resolution   & 0.8~kHz \\
Start Date             & 13 Mar 2024, 22:25 UTC \\
Observation Duration   & 4 hours \\
Time Resolution        & 1~s \\
Calibrator             & 3C196 \\
Target                 & Tau Boötis b \\
\hline
\end{tabular}
\end{table}

We pre-processed the raw data by flagging unwanted RFI, demixing, and averaging. For the flagging step, we used the \texttt{AOFlagger} strategy \citep{2010ascl.soft10017O}, which operates separately on each subband of around 200 kHz width. During the demixing step, we solved for instrumental gains in the directions of three dominant sources (Cassiopeia A, Cygnus A, and Virgo A) using models from \citet{de2020cassiopeia} and subtracted their contribution from the visibilities using these gains. Given that Virgo A is located less than $10^\circ$ from the phase centre, we used a smaller solution cadence of 10 s and 102.4 kHz to mitigate its effects. The decorrelation of the visibility phase within this solution interval is acceptable since the ionospheric conditions were relatively calm, as shown later.

After pre-processing, we processed the calibrator data using the Library for Low Frequencies \citep[LiLF;][]{de2018effect,de2019systematic}. The pipeline's latest version is robust and effectively handles rapid ionospheric variations and challenges posed by residual RFI. The calibration begins with an approximate bandpass correction, using the theoretical LBA bandpass response \citep{VanHaarlem2013}, followed by an additional RFI flagging step using \texttt{AOFlagger}'s `LBA wideband' strategy, which eliminates interference below 20~MHz.
Next, the pipeline corrects the polarisation misalignment, applies the beam correction, and compensates for differential Faraday rotation (details in \citealt{de2019systematic}). Diagonal phase solutions are then calculated. It is from these phase solutions that we must separate the contributions of clock and ionospheric delays. The pipeline then proceeds with bandpass calibration. However, this step is irrelevant to the current study as we are only concerned with the phase solutions hereafter.

\begin{figure*}
    \centering
    \includegraphics[width=1\linewidth]{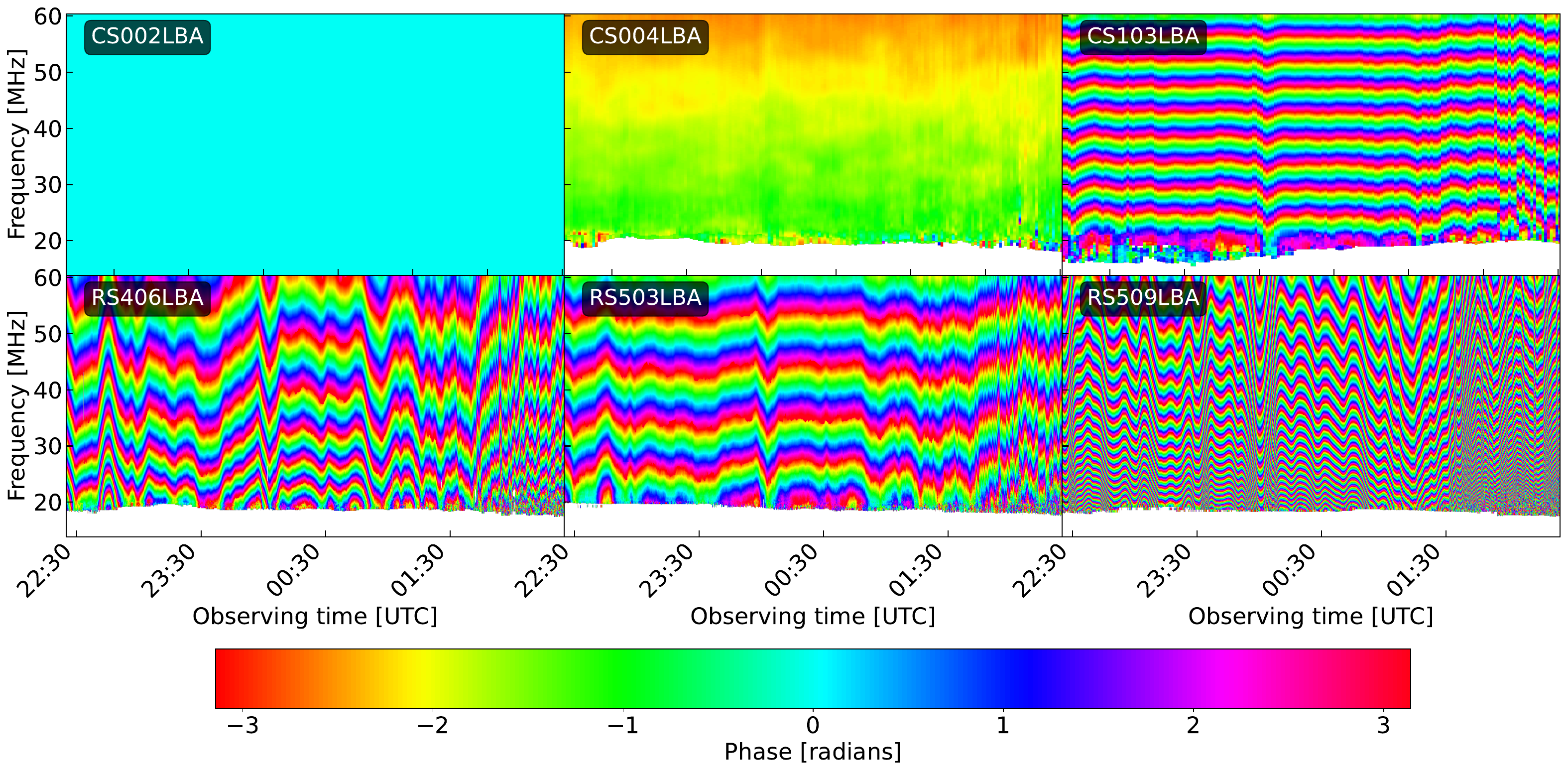}
    \caption{Phase solutions of the calibrator with station CS002 as reference. The frequencies below 20~MHz are RFI-corrupted and, therefore, flagged.}
    \label{fig:phases}
\end{figure*}

The calculated phase solutions are presented for a few example stations in Fig. \ref{fig:phases}, highlighting the relatively calm conditions during the observation, despite the observations occurring around solar maximum. The phase solutions reveal a known LOFAR software issue: CS103LBA shows large clock delays of several hundreds of nanoseconds.

\section{Method}\label{sec:method}

The algorithm presented in this paper can disentangle station delays from ionospheric effects using the noisy phase solutions from LOFAR LBAs. For simplicity, the algorithm is described in the context of a single interferometer element (station hereafter). The remaining stations follow an identical approach. All station phases are measured relative to a reference station because absolute station phases are meaningless in interferometry.

\subsection{Station phase model}\label{sec:model}

We modelled the observed phase as a sum of three terms: a frequency-independent phase offset, $\phi_0$, introduced by the electronics; a phase due to the station's clock delay, $\phi_C$;  and the phase due to dispersive propagation through the ionosphere's TEC\footnote{TEC is the same as column density.}, $\phi_{\rm{TEC}}$. The phase from a clock delay is proportional to frequency:
\begin{equation}
    \phi_{\rm C}(\Delta t,\nu) = 2\pi \Delta t \nu,
    \label{eq:phi_t}
\end{equation}where $\Delta t$ is the delay and $\nu$ is the frequency.

The ionospheric phase accumulated during propagation is given by an integral of the wavenumber along the line of sight \citep[LoS;][]{appleton1932wireless}:

\begin{equation}
    \phi_{\rm{TEC}}(\nu) = -\frac{2\pi\nu}{c} \int_{\rm LoS} (n-1) {\rm d}l,
    \label{eq:phi_tau_init}
\end{equation}
where $c$ is the speed of light, and $n$ is the ionospheric refractive index. Ignoring the influence of the Earth's magnetic field, the ionospheric plasma's refractive index is given by

\begin{equation}
    n = \sqrt{1-\left(\frac{\nu_p}{\nu}\right)^2}, 
    \label{eq:n}
\end{equation}
where the plasma frequency $\nu_{\rm p}$ is given by 
\begin{equation}
\nu_p = \frac{1}{2\pi} \sqrt{\frac{N_e e^2}{\epsilon_0 m_e}}.
\end{equation}
Here $e$ is the electron charge, $\epsilon_0$, the vacuum permittivity, $m_e$, the electron mass, and $N_e$ is the electron number density.

In Earth-based radio astronomy, even the lowest observed frequencies are significantly higher than the plasma frequency. In this regime, we applied a Taylor expansion of the refractive index and retained the first two terms, which largely determine the phase in our frequencies of interest. The first term (hereafter referred to as the first-order term) is dominant, while the second (the second-order term) typically becomes relevant only for frequencies below 40 MHz. The expression for the ionospheric phase then becomes

\begin{equation}
\begin{split}
    & \phi_{\rm{TEC}}(\tau_{1,2},\nu) \approx \frac{2\pi}{c} \left( \frac{a}{\nu} \tau_1 + \frac{a^2}{2\nu^3} \tau_2\right), \\
    & {\rm where}  \ \ \ a=10^{16}\frac{e^2}{8\pi^2 \epsilon_0 m_e} \ \ \ {\rm and} \ \ \ \tau_n = \int_{\rm LoS} N_e^n dl.
\end{split}
\label{eq:phi_tau_final}
\end{equation}
Here, $a$ is a constant and the $10^{16}$ factor arises due to the conversion from ${\rm electrons}/{\rm m}^2$ to ${\rm TEC}$  units. 

Gathering the three phase terms together, the final phase model that is fit to the data is

\begin{equation}
    \phi_{\rm M} \equiv \phi_{\rm M}(\phi_0,\Delta t, \tau_{1,2}, \nu) = \phi_0 + \phi_{\rm C}(\Delta t,\nu)+ \phi_{\rm{TEC}}(\tau_{1,2}, \nu),
    \label{eq:model}
\end{equation}
where $\phi_0$, $\Delta t$, $\tau_1$, and $\tau_2$ are model parameters to be estimated in the fitting process. The explicit form used in practice, with all constants evaluated, is
\begin{equation}
    \phi = \phi_0 + 2 \pi \Delta t \nu -8.4479745\cdot 10^{9}\frac{\tau_1}{\nu} +  10^{21} \frac{\tau_2}{\nu^3},
    \label{eq:losoto}
\end{equation}
where the physical constants have been absorbed into the numerical coefficients preceding each term.

\subsection{Likelihood definition}

In an ideal, noise-free scenario, the model phases, $\phi_{\rm M}$ will equal the observed phases $\phi_{\rm D}$. In the presence of noise, we wish to minimise the difference between $\phi_{\rm M}$ and $\phi_{\rm D}$. A brute-force approach to achieving this would be to evaluate the squared difference on a four-dimensional grid of all plausible parameter values, $\phi_0,\Delta t$, and $\tau_{1,2}$, and locate the minimum. However, this method is computationally expensive. Instead, we solved for the model parameters that maximise the likelihood, which is the probability of observing the data given the model parameters.

Since the noise in the measured phases at different channels is uncorrelated, the likelihood ($L$) is the product of the likelihood of individual channels:

\begin{equation}
L = \prod_{i=1}^{N_{\rm ch}} f(\phi_{{\rm D},i}, \phi_{{\rm M},i}, \kappa),
\label{eq:likelihood}
\end{equation}
 where $N_{\rm ch}$ is the number of frequency channels, $i$ is the channel index, and $f(\cdot)$ is the likelihood for a single channel. Because the phase is an angular quantity, we modelled it using the von Mises distribution, with the model prediction $\phi_{\rm M}$ acting as the mean. That is,

\begin{equation}
f(\phi_{{\rm D},i}, \phi_{{\rm M},i}, \kappa) = \frac{1}{2\pi I_0(\kappa)} e^{\kappa \cos(\phi_{{\rm D},i} - \phi_{{\rm M},i})},
\end{equation}where $\kappa$ is the so-called concentration, and $I_0$ is the modified Bessel function of order zero. The concentration parameter is a measure of dispersion due to noise and is analogous to the inverse variance of a normal distribution.

\subsection{Estimating the concentration parameter}
\label{subsec:concentration}
The calculation of the likelihood requires us to estimate $\kappa$ from the data, which is not straightforward. Consider samples, $\phi_i$, $i\in[0,N_{\rm ch})$, drawn from a von Mises distribution with zero mean and some concentration, $\kappa$. In our case, these samples represent the difference between the measured phases and the model with the `true' parameter values. In the high signal-to-noise-ratio limit and in the absence of phase-wrapping, $\kappa$ can be estimated as the inverse of the sample variance of the phases $\phi_i$. Neither of these conditions is true in our case. An alternative route to estimating $\kappa$ is to first estimate the so-called mean resultant length (also called the circular mean) given by

\begin{equation}
    r = \left| \frac{1}{N_{\rm ch}}\sum \exp({\rm i}\phi_i) \right|.
\end{equation}
If the data have a high amount of dispersion (low concentration) then the resultant length tends to zero, and in the opposite case, the length tends to unity. 

The maximum likelihood estimate of $\kappa$ given the resultant length can be shown to be \citep{marrelec2024estimating}

\begin{equation}
\begin{split}
    & r = \frac{I_1(\kappa)}{I_0(\kappa)}, \\ 
    &  {\rm where} \ \ \ I_0(\kappa) = \frac{1}{\pi} \int_0^\pi e^{\kappa \cos(\phi)} d\phi \\
    & {\rm and}  \ \ \ I_1(\kappa) = \frac{1}{\pi} \int_0^\pi e^{\kappa \cos(\phi)} \cos(\phi)d\phi.
\end{split}
\label{eq:r_th}
\end{equation}
Here, $I_0$ and $I_1$ are the modified Bessel functions of order zero and one, respectively. There is no known closed form solution for $\kappa$ given $r$, so Eq. \ref{eq:r_th} must be solved iteratively. 

An initial approximation of $\kappa$ ($\kappa_0$) can be obtained using \citep{mardia2009directional}

\begin{equation}
    \kappa_0 = \frac{r(2-r^2)}{1-r}
    \label{eq:kappa0}.
\end{equation}We further refined the value of $\kappa$ using Newton's method to find the roots of 
\begin{equation}
    g(\kappa_i) = \frac{I_1(\kappa)}{I_0(\kappa)} - r = 0.
    \label{eq:g}
\end{equation}The iterations are given by 
\begin{equation}
    \kappa_{i+1} = \kappa_{i} - \frac{g(\kappa_i)}{g'(\kappa_i)}.
    \label{eq:kappa_1}
\end{equation}The derivative can be written out analytically (details in Appendix \ref{sec:derivation}) and reads
\begin{equation}
    \kappa_{i+1}= \kappa_i - \left[\frac{I_1(\kappa_i)}{I_0(\kappa_i)} - r\right]\left[1-\left(\frac{I_1(\kappa_0)}{I_0(\kappa_0)} \right)^2 - \frac{r}{\kappa_0}\frac{I_1(\kappa_i)}{I_0(\kappa_i)}\right]^{-1}
.\end{equation}To achieve the desired precision of $10^{-3}$, we performed two maximum likelihood estimation refinement iterations, reusing  Eqs. \ref{eq:g} and \ref{eq:kappa_1}.

\subsection{Parameter estimation}
 
Before starting the fitting, we filled in missing flagged values for the observed phases in the phasor space via linear interpolation of the real and imaginary components. This approach is effective provided that only a few consecutive points are flagged, which is typically the case for our data. If more than 60\% of the frequencies in a time slot are flagged, the clock delays and ionospheric TEC will be flagged and interpolated later.

To avoid numerical precision issues, we maximised the log of the likelihood from  Eq. \ref{eq:likelihood}:

\begin{equation}
    \log L = \sum_{i=1}^{N_{\rm ch}} \log f(\phi\rm{_i}; \mu, \kappa).
\label{eq:loglikelihood}
\end{equation}We calculated the log-likelihood using the function \texttt{scipy.stats.vonmises.logpdf} in the \texttt{scipy} package.
   
Since the model is non-linear in its parameters, we used the gradient-based optimisation algorithms \texttt{fmin\_bfgs} implemented in \texttt{scipy}. \citep{virtanen2020scipy}. The Broyden-Fletcher-Goldfarb-Shanno (BFGS) algorithm is a quasi-Newton optimisation method that approximates the inverse Hessian matrix to guide its search for the optimal parameters. Starting from an initial guess, the algorithm iteratively refines the parameters by evaluating the gradient of the objective function to determine the search direction and step size. Unlike limited-memory variants, \texttt{fmin\_bfgs} computes and stores the full inverse Hessian approximation, making it well suited for problems of moderate dimensionality.  \texttt{fmin\_bfgs} is particularly effective for optimisation problems where gradients can be calculated efficiently. Since this algorithm minimises the objective function by default, we adapted it to maximise the log-likelihood by minimising the negative log-likelihood function instead.

Several challenges arise when estimating the parameters. Firstly, we found that accurate approximations of the initial values are essential for ensuring convergence to the correct solution. Secondly, phase offsets and delays exhibit significant degeneracy, complicating their precise determination. Finally, in order to calculate the best-fitting model parameters by maximising likelihood, we must know the concentration parameter. However, to know the concentration parameter via the procedure in Sect. \ref{subsec:concentration}, we need to know the best-fit parameters in order to estimate the level of noise in the data. 

To solve the problem of estimating the model parameters and the concentration, we adopted a two-step approach. In the first step, we estimated the concentration parameter directly from the observed phases. Although the phases contain both signal and noise -- rather than purely noise, as would be ideal for estimating concentration -- this approach yields a sufficiently accurate first-order approximation. Next, we obtained initial estimates for the delay, the two ionospheric terms, and phase offset. The delay, $\Delta t$, was estimated by performing a Fourier transform of the data, where any phase slope caused by delay manifests as a peak in the Fourier space. Starting from this delay estimate, we estimated the phase offset, $\phi_0$, and the first-order ionospheric term via a brute-force grid search. Specifically, we varied the phase offset between $-\pi$ and $\pi$, the delay between $\pm 20$~ns of the Fourier estimate for core stations, $\pm 200$~ns for remote stations, and $\pm 500$~ns for international stations. The first-order ionospheric term, given by the TEC, was varied between $-0.05$~TECU and $0.05$~TECU for core stations, and $-1.5$~TECU and $1.5$~TECU for remote stations. In this iteration, we neglected the second-order effects of the ionosphere for simplicity. Additionally, we solved for parameters across multiple time slots simultaneously, assuming the phase offset remains constant across all of them while allowing the delay and ionospheric term to vary. This assumption provides a robust initial approximation for the fitting parameters, $\phi_0$, $\Delta t$, and $\tau_1$.

Before proceeding to the second iteration, we smoothed the initial solutions with a median filter (\texttt{scipy.signal.medfilt} with the kernel size 5). Using the initial parameters, we calculated the residual phases (i.e. data minus model) that must be largely dominated by noise. From the residual phases, we calculated the improved concentration parameter using the procedure in Sect. \ref{subsec:concentration}. Armed with this estimate of the concentration parameter, we could find the best-fit parameters, including the second-order ionospheric terms, by maximising the log-likelihood.

A key aspect of this procedure is how we handle parameter estimation across different time steps. While it is possible to fit each time step independently, the temporal smoothness of the physical parameters allows for more stable solutions when leveraging continuity in time. For the first time slot, we used the parameters obtained in the previous step as initial guesses. Additionally, if the lowest observed frequency is below 40~MHz, we included the second-order ionospheric term in the model, with its initial value estimated via a brute-force grid search.

Once the parameters for the first time slot were determined, we quantified the fit quality using the circular standard deviation:

\begin{equation}
    \sigma_{\text{circ}} = \sqrt{-2{\rm ln}(r)}, \ \ \ {\rm where}  \ \ \  r =  \left| \frac{\sum_{i=1}^{N_{\rm ch}} \kappa\rm{_i} \theta_{{\rm R}_i}}{\sum_{i=1}^{N_{\rm ch}} \kappa\rm{_i}} \right|
    \label{eq:std_circ}
.\end{equation}Here, $\theta_{{\rm R}_i}$ are the residual phase phasors and $\kappa\rm{_i}$ are their associated concentration parameters. The statistic $\sigma_{\text{circ}}$ provides a measure of the noise level in the fit.

For subsequent time slots, we used the estimated parameters from the previous slot as initial guesses, exploiting the typically smooth evolution of the delay and ionospheric terms over time. These parameters are then refined using the \texttt{fmin\_bfgs} optimisation algorithm. If the resulting circular standard deviation remains comparable to that of the initial fit, the updated parameters are accepted. If the deviation increases significantly, the initial parameters are re-estimated using a brute-force grid search.

Although this second iteration yields reliable results in most cases, occasional outliers due to RFI or data noise may still occur. Since station delays and ionospheric effects are expected to vary smoothly, we applied a final post-processing step: a median filter (\texttt{scipy.signal.medfilt}, kernel size 5) to suppress any residual anomalies.

The method described above is implemented in  \texttt{Python} \footnote{The complete implementation is available at \url{https://git.astron.nl/cordun/clocktec_loglik}}. It has been validated for the LBA system. The HBA system is not currently included, as the standard processing pipeline performs adequately in that regime. Nonetheless, adapting this method to HBAs remains a potential direction for future development, particularly in the context of LOFAR 2.0.

\section{Results} \label{sec:results}

\begin{figure}
    \centering
    \includegraphics[width=1\linewidth]{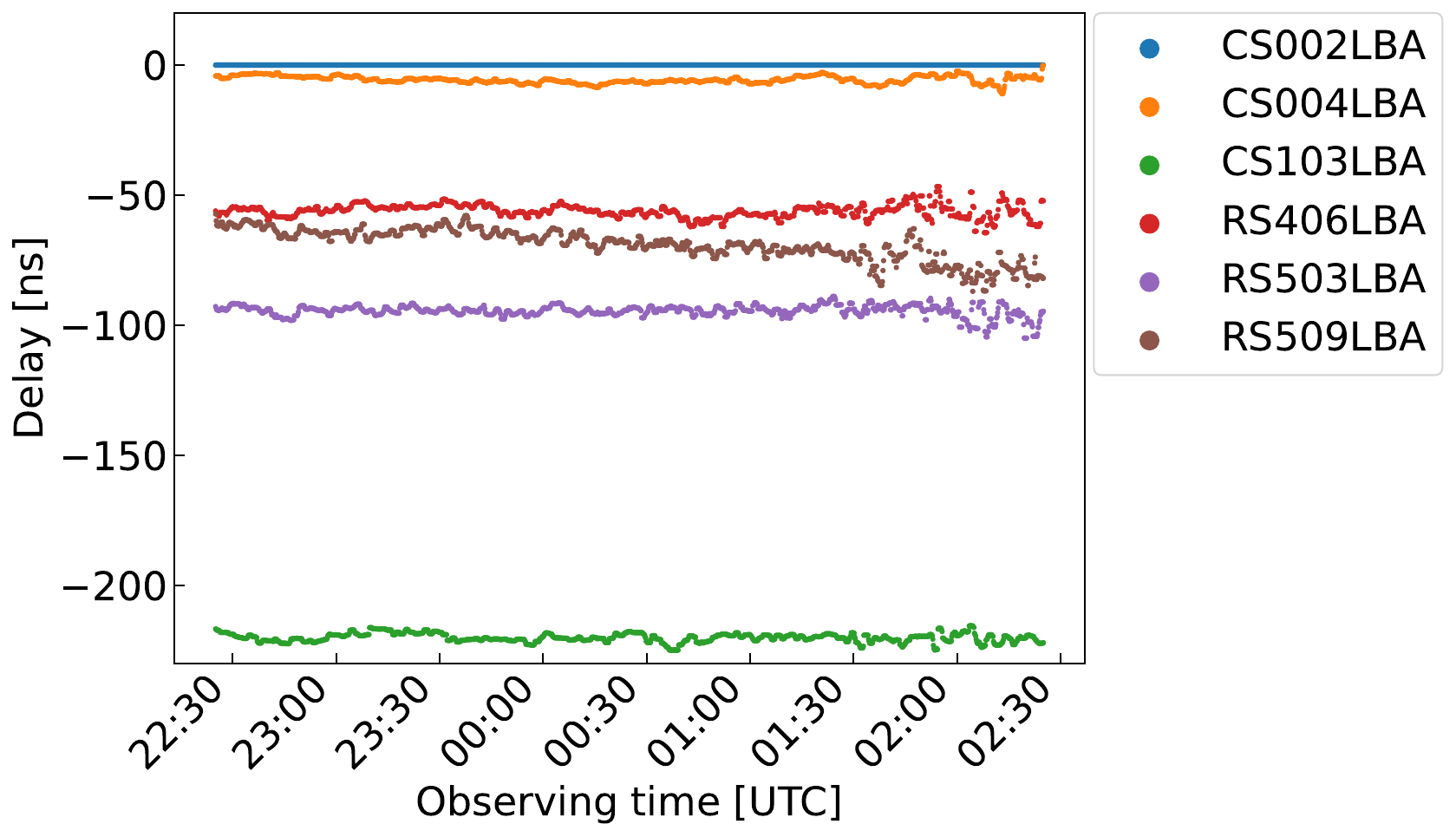}
    \caption{Delays between stations for the calibrator with station CS002 as reference.}
    \label{fig:clock}
\end{figure}

\begin{figure}
    \centering
    \includegraphics[width=1\linewidth]{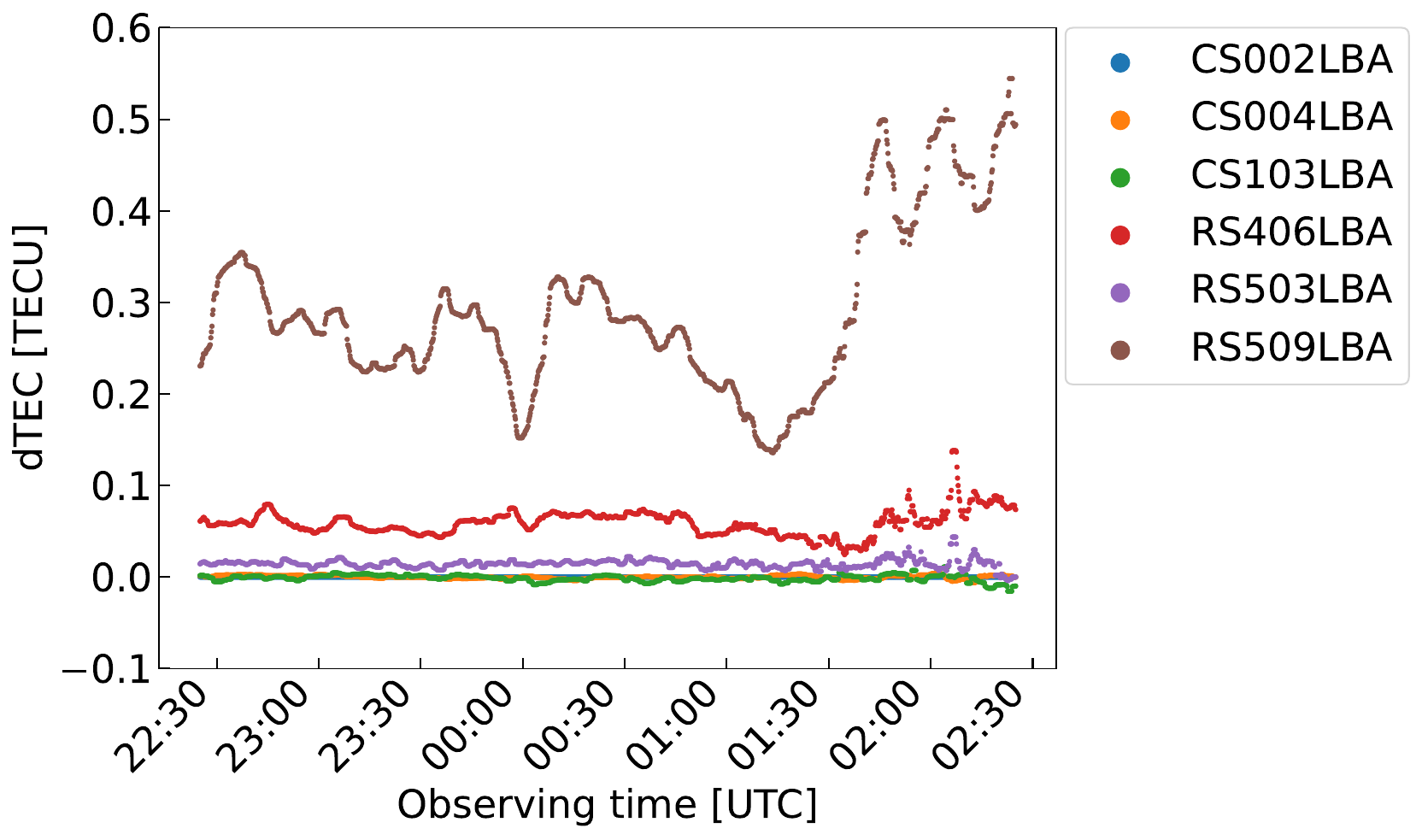}
    \caption{First-order ionospheric term for the calibrator with station CS002 as reference. CS002 (blue) dots are hidden behind the CS004 (orange) dots because both TEC values are approximately 0.}
    \label{fig:tec1}
\end{figure}

\begin{figure}
    \centering
    \includegraphics[width=1\linewidth]{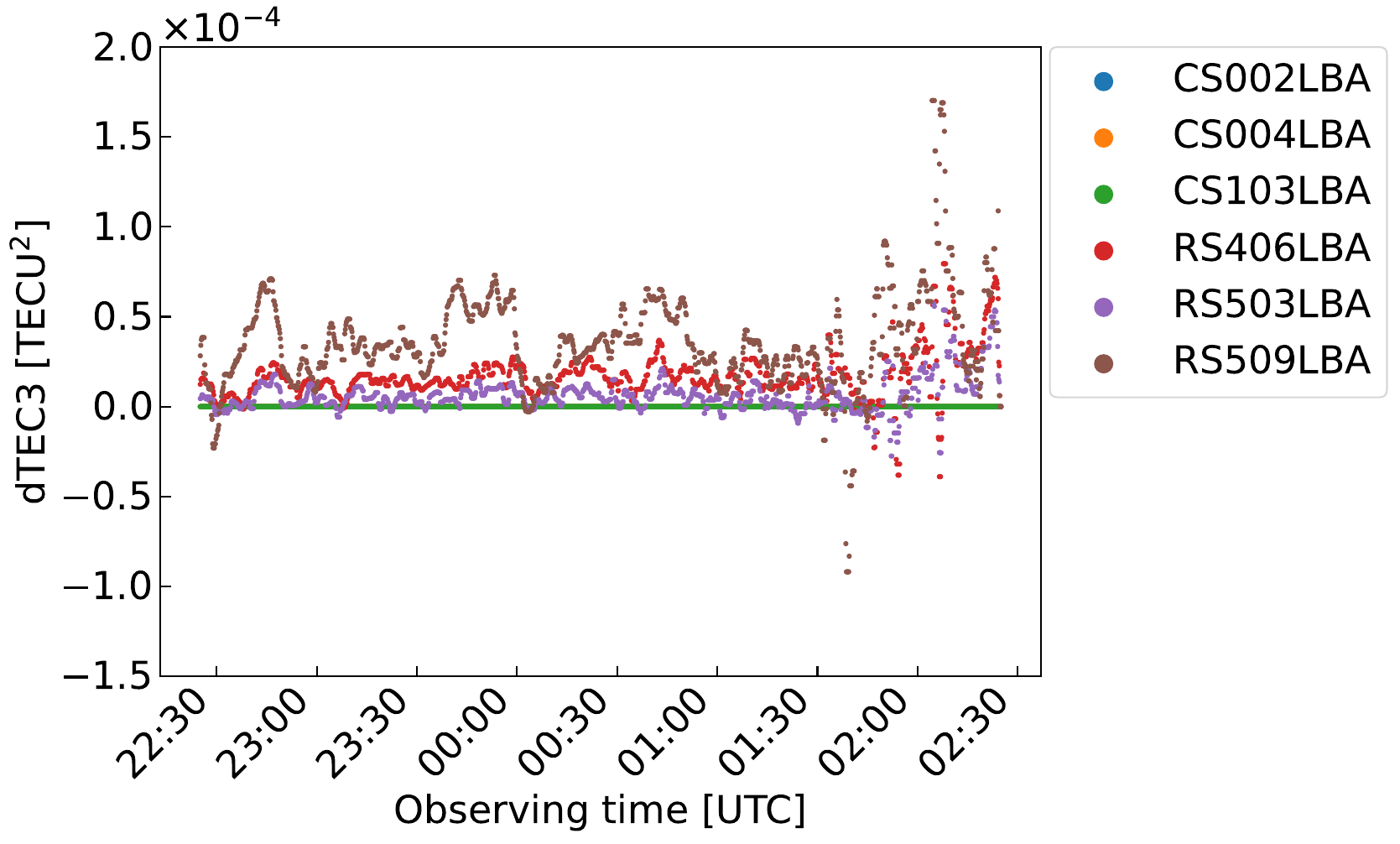}
    \caption{Second-order ionospheric term for the calibrator with station CS002 as reference. All the core stations have a second-order ionospheric term of approximately 0.}
    \label{fig:tec3}
\end{figure}

The fitting function produces three key outputs: the station delays (Fig. \ref{fig:clock}), the first-order ionospheric term (Fig. \ref{fig:tec1}), and the second-order ionospheric term (Fig. \ref{fig:tec3}). In the plots, we do not show all stations, just a few representative examples. As expected, the differences between the core stations are small because they are located close together and are synchronised to the same clock. The only exception is CS103, which shows significant delays relative to the reference station. These are due to a known software issue.

Over the observation period, the parameters generally change smoothly over time, with occasional outliers caused by RFI. Towards the end of the observation, the solutions become noisier. This increase in noise occurs because the calibrator is positioned at a low elevation in the sky, where strong atmospheric effects and reduced signal strength impact the quality of the measurements. Despite the increased noise, the solutions remain reliable, as they still follow the expected trends and do not show discontinuities or non-physical behaviour.

\section{Discussion}\label{sec:disc}

\begin{figure*}
    \centering
    \includegraphics[width=1\linewidth]{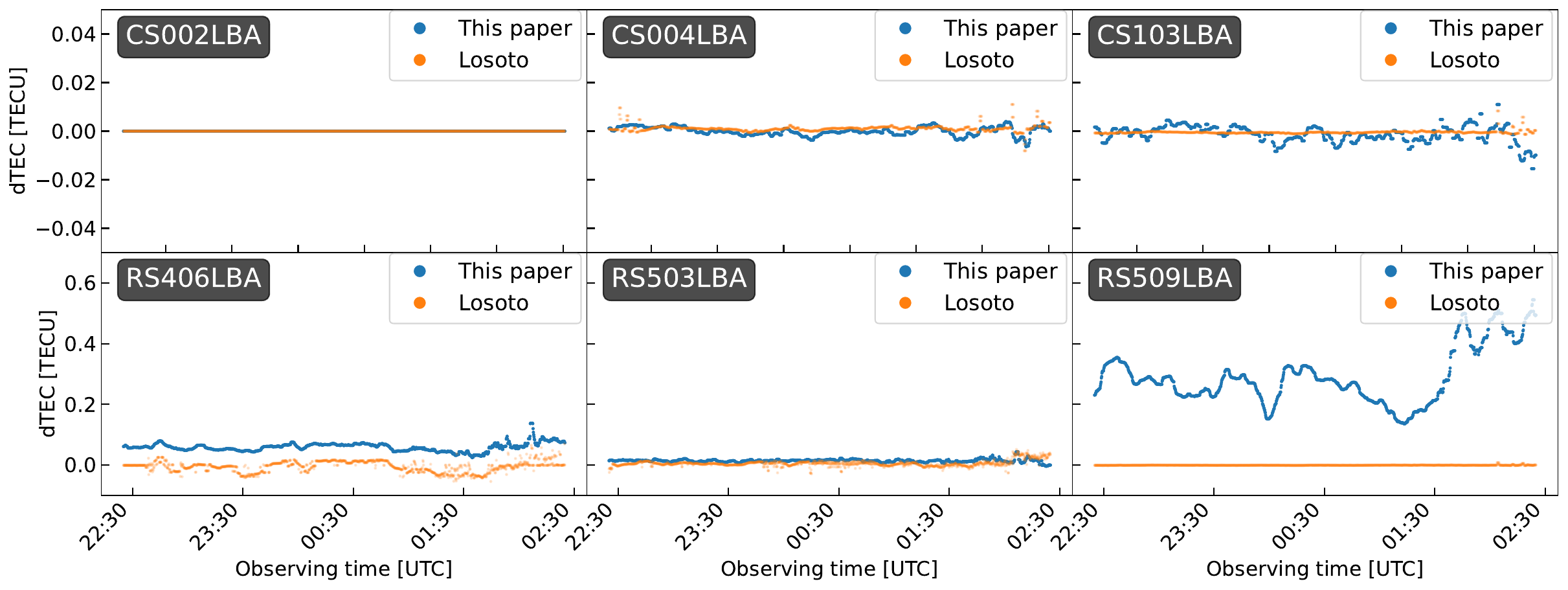}
    \caption{Comparison of the ionospheric first-order terms derived in this paper and those computed using \texttt{losoto} for the example stations.}
    \label{fig:comp_tec}
\end{figure*}

\begin{figure*}
    \centering
    \includegraphics[width=1\linewidth]{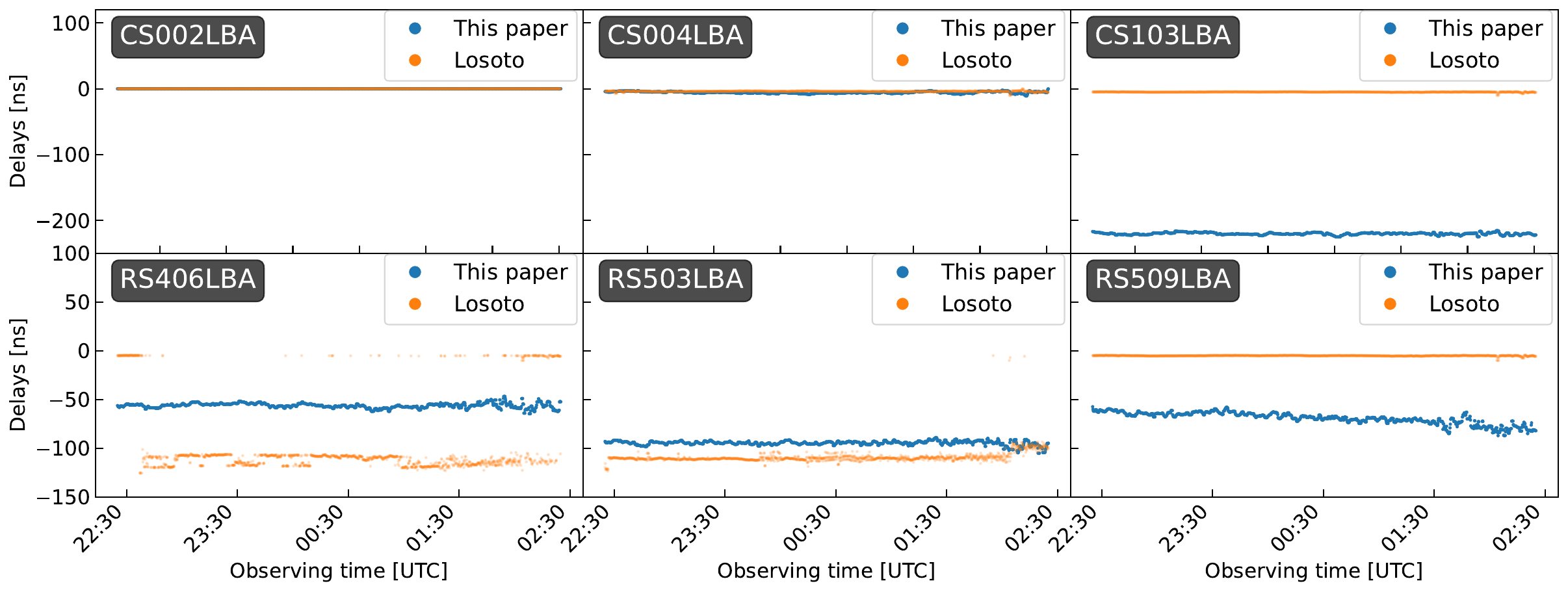}
    \caption{Comparison of the station delays derived in this paper and those computed using \texttt{losoto} for the example stations.}
    \label{fig:comp_delay}
\end{figure*}

For comparison with the state-of-the-art, we ran the clock-TEC separation algorithm implemented in \texttt{losoto} \citep{de2019systematic} on the same phase solutions using the same reference station. The differences between the resulting ionospheric first-order terms and the station delays are shown in Figs. \ref{fig:comp_tec} and \ref{fig:comp_delay}. During the process, the \texttt{losoto} code returned several warnings about large $\chi^2$ values, suggesting that the fit might not be accurate. While the algorithm produced reasonable estimates for some stations (e.g. RS503LBA and CS004LBA), its performance was inconsistent across the array. In several cases, the algorithm failed in one of two distinct ways: either it returned a value of zero, indicating no valid fit was found (e.g. CS103LBA and RS509LBA), or it returned an incorrect estimate that visibly deviates from the expected trend (e.g. RS406LBA).

\begin{figure*}[h!t]
    \centering
    \begin{subfigure}[t]{0.4\textwidth}
        \centering
        \includegraphics[width = 1\linewidth]{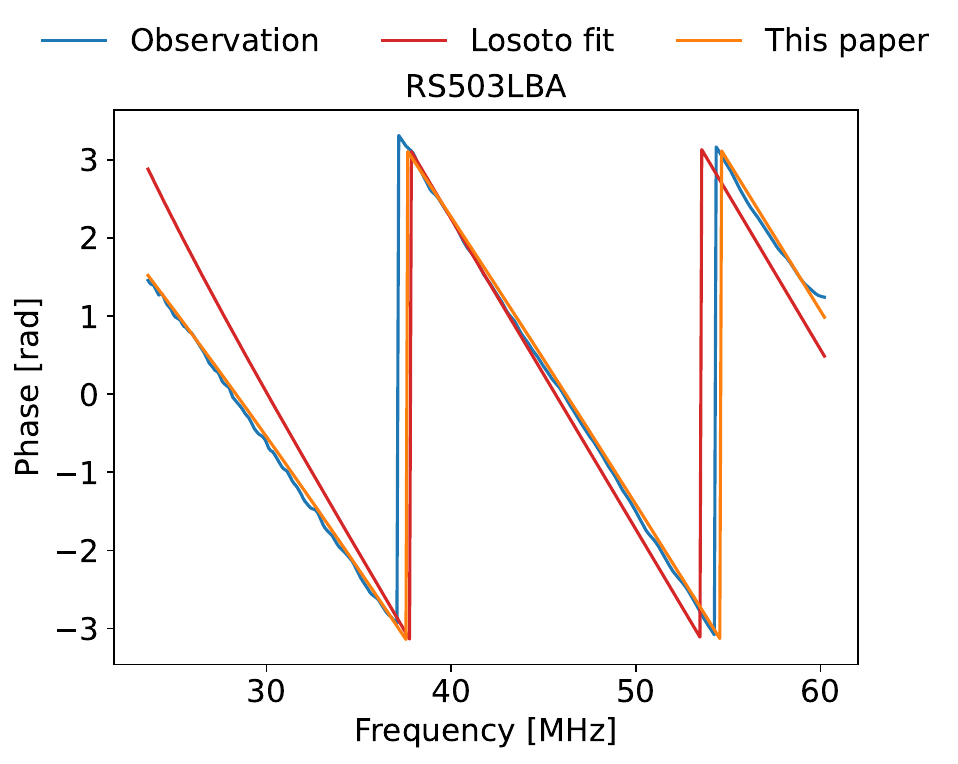}
    \end{subfigure}%
    ~ 
    \begin{subfigure}[t]{0.4\textwidth}
        \centering
        \includegraphics[width = 1\linewidth]{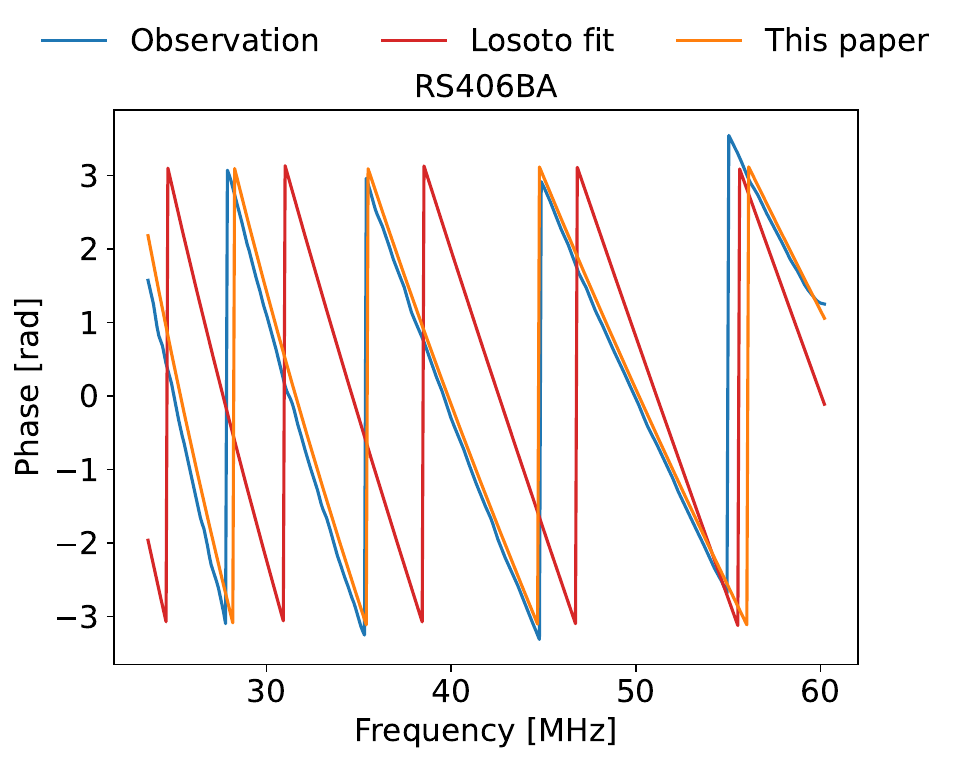}
    \end{subfigure}

    \caption{Comparison of the observed phases, the fits obtained with this algorithm and \texttt{losoto}. There are two cases: one where the fitting parameters of the two codes match (left) and one where they do not (right).}
    \label{fig:comp_fit}
\end{figure*}

To assess which method provided a more reliable estimate, we compared the fitted parameters from both methods against the observed phase data. For \texttt{losoto}, we constructed the model using the parameters returned by its fitting routine presented in Eq. \ref{eq:losoto}. We show examples of matching results and mismatches in Fig. \ref{fig:comp_fit}. For RS406LBA (mismatch - left), \texttt{losoto} produces incorrect parameter estimates, unlike RS503LBA (match - right), where the results are better, but there is still a difference towards the lowest frequencies due to a wrong estimation of the second-order ionospheric term.

The main difference lies in how the phase offset is treated. In \texttt{losoto}, designed for HBAs, the phase offset is nearly zero and is seamlessly integrated into the first ionospheric term due to the narrow relative bandwidth. This narrow bandwidth also introduces significant degeneracies in HBAs, resolved by effectively ignoring the phase offset. However, for LBAs, the larger relative bandwidth causes stronger ionospheric contributions and fewer degeneracies. In this case, the phase offset is not zero and cannot be ignored, making \texttt{losoto} less effective.

While the improved clock--TEC separation method presented in this paper is more accurate, it comes with a trade-off: higher computational cost. Unlike the \texttt{losoto} code, which uses a linear fit, our method minimises the likelihood to determine the parameters, requiring more processing power.

Despite this drawback, the new method is much more reliable and robust. It successfully resolves delay and ionospheric solutions for datasets from solar maximum conditions over a wide frequency range. The algorithm can handle various ionospheric conditions, failing only during extreme conditions, such as geomagnetic storms, where calibration is impractical regardless.

Looking ahead, the primary objective is to integrate this method with the \texttt{losoto} code to develop a unified algorithm\footnote{The new algorithm can be found at \url{https://git.astron.nl/mevius/clocktec}}. This new approach aims to combine the robustness of the current method with the computational efficiency of \texttt{losoto}. The new algorithm will use von Mises concentration-based weights and a log-likelihood method to generate reliable initial solutions, which will then facilitate phase unwrapping. Once unwrapped, a linear fit will be applied to extract key parameters: a constant phase offset in time, as well as station delays and ionospheric terms that vary over time. By merging these techniques, the resulting method will be both accurate and efficient, balancing reliability and speed for improved calibration at low frequencies.

Currently, the LOFAR LBA survey requires at least two simultaneous beams: one on the target and one on the calibrator. This is because delay and ionospheric effects cannot yet be reliably separated, requiring the complete phase solutions from the calibrator to be applied to the target. However, an improved algorithm is a step towards more efficient, large-scale LOFAR LBA surveys, where continuous calibrator observation may no longer be necessary.

\section{Conclusion}\label{sec:conclusion}

With low-frequency radio astronomy it is challenging to separate station delays from ionospheric effects, particularly at ultra-low frequencies such as those observed with LOFAR’s LBAs. Existing methods, such as \texttt{losoto}, often struggle in this regime, leading to unreliable or incomplete solutions. This is largely because \texttt{losoto} was developed for higher-frequency observations, where the relative bandwidth is narrower, simplifying the separation process. At LBA frequencies, however, the larger relative bandwidth amplifies ionospheric contributions and reduces degeneracies, demanding more robust and specialised algorithms for accurate calibration.

\onecolumn

\twocolumn

In this paper, we have presented a new clock--TEC separation algorithm that replaces traditional phase unwrapping and linear fitting with a likelihood-based approach using the von Mises distribution. This approach significantly improves the reliability and accuracy of delay and ionospheric solutions, even in demanding scenarios such as at solar maximum or when the relative bandwidth is large

The proposed method achieves robust clock--TEC separation. By effectively handling noisy and flagged data, the algorithm ensures consistent results, reducing the dependence on simultaneous calibrator observations. This advancement not only enhances LOFAR’s ionospheric monitoring capabilities but also provides a strong foundation for improving the efficiency of large-scale, all-sky surveys.

Even though LOFAR 2.0 will synchronise the clocks for the Dutch stations, the international stations will still experience significant delays. The algorithm presented in this work could help address these delays, ensuring more accurate calibration across the array. By improving the separation of the clock and ionospheric effects, this method can fully maximise LOFAR’s observational time. Specifically, it removes the need to observe the calibrator and target simultaneously, allowing more beams to be allocated to the target fields after a separate calibration scan.

\begin{acknowledgements}

CMC and HKV acknowledge funding from the European Research Council via the starting grant `STORMCHASER' (grant number 101042416). CMC thanks Dr. Henrik Edler for the insightful discussions. This work made use of the Python packages \texttt{matplotlib} \citep{mpl} to generate the figures and \texttt{numpy} \citep{np} for computations. 
\end{acknowledgements}

%
%

\bibliographystyle{aa}
\bibliography{bibliography}

\begin{appendix}

\onecolumn
\section{Iterative solution for $\kappa$} \label{sec:derivation}

In this section we derive Eq. \ref{eq:kappa_1}, restated here:

\begin{equation}
    \kappa_{1} = \kappa_0 - \frac{g(\kappa_0)}{g'(\kappa_0)} 
    \label{eq:kappa_1_2}.
\end{equation}We defined the function $g(\kappa_0)$ that will be minimised,

\begin{equation}
    g(\kappa_0) = r_{\rm th}(\kappa_0) - r_{\rm approx}.
    \label{eq:g_2}
\end{equation}We found $r_{\rm{approx}}$ using

\begin{equation}
    r_{\rm approx} = \frac{1}{N_{\rm ch}}\sum_0^{N_{\rm ch}} \exp\left({\rm i} \phi_{\rm D}\right),
\label{eq:r_approx2}
\end{equation}and $r_{\rm{th}}(\kappa_0)$ using 

\begin{equation}
    r_{\rm th}(\kappa) = \frac{I_1(\kappa)}{I_0(\kappa)},
\label{eq:r_th_3}
\end{equation}where $I_0(\kappa_0)$ and $I_1(\kappa_0)$ are the modified Bessel functions, which are

\begin{equation}
    \begin{split}
        & I_0(\kappa) = \frac{1}{\pi} \int_0^\pi e^{\kappa \cos(\phi)} d\phi \\
        & I_1(\kappa) = \frac{1}{\pi} \int_0^\pi e^{\kappa \cos(\phi)} \cos(\phi)d\phi.
    \end{split}
    \label{eq:bessel_2}
\end{equation}Next, we calculated the derivative, 

\begin{equation}
    g'(\kappa_0) = \left( \frac{I_1(\kappa_0)}{I_0(\kappa_0)} -  r_{\rm approx}\right)' = \left( \frac{I_1(\kappa_0)}{I_0(\kappa_0)}\right)',
    \label{eq:deriv1}
\end{equation}and we used the quotient rule for derivatives to obtain

\begin{equation}
    g'(\kappa_0) = \frac{I_1(\kappa_0)' I_0(\kappa_0) - I_1(\kappa_0) I_0(\kappa_0)'}{I_0(\kappa_0)^2}.
\label{eq:deriv2}
\end{equation}The derivative of $I_0(\kappa_0)$ was directly calculated, 

\begin{equation}
    I_0(\kappa_0)' = \frac{1}{\pi} \int_0^\pi e^{\kappa \cos(\phi)} \cos(\phi)d\phi = I_1(\kappa_0),
    \label{eq:deriv_i0}
\end{equation}

\noindent while the derivative of $I_1(\kappa_0)$ was obtained from the recurrence formula and is

\begin{equation}
    I_1(\kappa_0)' = I_0(\kappa_0) - \kappa_0 I_1(\kappa_0)
    \label{eq:deriv_i1}.
\end{equation}Substituting these derivatives into Eq. \ref{eq:deriv2}, we obtain  

\begin{equation}
    g'(\kappa_0) = \frac{(I_0(\kappa_0) - \frac{1}{\kappa_0} I_1(\kappa_0)) I_0(\kappa_0) - I_1(\kappa_0) I_1(\kappa_0)}{I_0(\kappa_0)^2},
    \label{eq:deriv3}
\end{equation}which simplifies to 

\begin{equation}
    g'(\kappa_0) = 1- r_{\rm th}(\kappa_0)^2 - \frac{1}{\kappa_0}r_{\rm th}(\kappa_0). 
    \label{eq:deriv4}
\end{equation}Finally, we substituted Eq. \ref{eq:deriv4} into Eq. \ref{eq:kappa_1_2}, obtaining

\begin{equation}
    \kappa_{1} = \kappa_0 - \frac{r_{\rm th}(\kappa_0) - r_{\rm approx}}{1-r_{\rm th}(\kappa_0)^2 - \frac{1}{\kappa_0} r_{\rm th}(\kappa_0)}.
    \label{eq:kappa_1_3}
\end{equation}

\end{appendix}
\end{document}